\documentclass[aps,prb,twocolumn,reprint,showpacs,superscriptaddress,floatfix]{revtex4-1}
\usepackage{graphicx, epsfig}
\usepackage{color}
\begin{document}

\preprint{ }
\title{Electronic structure of an antiferromagnetic metal: CaCrO$_3$}

\author{P. A. Bhobe}
\email{preeti@spring8.or.jp}
\affiliation{Institute for Solid State Physics, The University of Tokyo, Kashiwa, Chiba 277-8581, Japan}
\affiliation{RIKEN, SPring-8 Centre, Sayo-cho, Sayo-gun, Hyogo 679-5148, Japan}
\author{A. Chainani}
\affiliation{RIKEN, SPring-8 Centre, Sayo-cho, Sayo-gun, Hyogo 679-5148, Japan}
\author{M. Taguchi}
\affiliation{RIKEN, SPring-8 Centre, Sayo-cho, Sayo-gun, Hyogo 679-5148, Japan}
\author{R. Eguchi}
\affiliation{Institute for Solid State Physics, The University of Tokyo, Kashiwa, Chiba 277-8581, Japan}
\affiliation{RIKEN, SPring-8 Centre, Sayo-cho, Sayo-gun, Hyogo 679-5148, Japan}
\author{M. Matsunami}
\affiliation{Institute for Solid State Physics, The University of Tokyo, Kashiwa, Chiba 277-8581, Japan}
\affiliation{RIKEN, SPring-8 Centre, Sayo-cho, Sayo-gun, Hyogo 679-5148, Japan}
\author{T. Ohtsuki}
\affiliation{RIKEN, SPring-8 Centre, Sayo-cho, Sayo-gun, Hyogo 679-5148, Japan}
\author{K. Ishizaka}
\affiliation{Institute for Solid State Physics, The University of Tokyo, Kashiwa, Chiba 277-8581, Japan}
\author{M. Okawa}
\affiliation{Institute for Solid State Physics, The University of Tokyo, Kashiwa, Chiba 277-8581, Japan}
\author{M. Oura}
\affiliation{RIKEN, SPring-8 Centre, Sayo-cho, Sayo-gun, Hyogo 679-5148, Japan}
\author{Y. Senba}
\affiliation{JASRI/SPring-8, Sayo-cho, Sayo-gun, Hyogo 679-5198, Japan}
\author{H. Ohashi}
\affiliation{JASRI/SPring-8, Sayo-cho, Sayo-gun, Hyogo 679-5198, Japan}
\author{M. Isobe}
\affiliation{Institute for Solid State Physics, The University of Tokyo, Kashiwa, Chiba 277-8581, Japan}
\author{Y. Ueda}
\affiliation{Institute for Solid State Physics, The University of Tokyo, Kashiwa, Chiba 277-8581, Japan}
\author{S. Shin}
\affiliation{Institute for Solid State Physics, The University of Tokyo, Kashiwa, Chiba 277-8581, Japan}
\affiliation{RIKEN, SPring-8 Centre, Sayo-cho, Sayo-gun, Hyogo 679-5148, Japan}

\date{\today}

\begin{abstract}
We report on the electronic structure of the perovskite oxide CaCrO$_3$ using valence-band, core-level, and Cr $2p-3d$ resonant photoemission spectroscopy (PES). Despite its antiferromagnetic order, a clear Fermi edge characteristic of a metal with dominant Cr $3d$ character is observed in the valence band spectrum. The Cr $3d$ single particle density of states are spread over 2 eV, with the photoemission spectral weight distributed in two peaks centered at $\sim$ 1.2 eV and 0.2 eV below E$_F$, suggestive of the coherent and incoherent states resulting from strong electron-electron correlations. Resonant PES across the Cr $2p-3d$ threshold identifies a `two-hole' correlation satellite and yields an on-site Coulomb energy $U\sim$ 4.8 eV. The metallic DOS at E$_F$ is also reflected through the presence of a well-screened feature at low binding energy side of the Cr $2p$ core-level spectrum. X-ray absorption spectroscopy (XAS) at Cr L$_{3,2}$ and O $K$ edges exhibit small temperature dependent changes that point towards a small change in Cr-O hybridization. The multiplet splitting in Cr $2p$ core level spectrum as well as the spectral shape of the Cr XAS can be reproduced using cluster model calculations which favour a negative value for charge transfer energy between the Cr $3d$ and O $2p$ states. The overall results indicate that CaCrO$_3$ is a strongly hybridized antiferromagnetic metal, lying in the regime intermediate to Mott-Hubbard and charge-transfer systems.
\end{abstract}
\maketitle

\section*{Introduction}
The origin of band gaps and character of the valence electrons in $3d$ transition metal compounds have been
central to the description of correlated electron systems. Transition metal oxides have fascinated the field of
condensed matter physics with their exotic physical properties, such as high-Tc superconductivity in cuprates
\cite{bed}, one-dimensional charge and spin self-organization in nickelates \cite{tran} and charge and orbital
ordering in manganites \cite{tok}. Numerous efforts to elucidate the underlying physics led to a broad
classification scheme of such materials into two categories viz. the Mott-Hubbard type and the charge-transfer
type \cite{zaa}. The compounds of early transition metals such as Ti, V are thought to be Mott-Hubbard type,
while the compounds of late transition metals like Ni, Cu form the charge-transfer type. The basis for such a
classification stands on the relative value of the Coulomb interaction energy $U$ between the $d$ electrons and
the one electron band width $W$. In the limit of large $U$, this system of singly occupied sites can be described
by the antiferromagnetic Heisenberg spin Hamiltonian. The opposite limit of large $W$, on the other hand, gives
an uncorrelated half filled metallic band. In other words, a system becomes metallic when $U/W <$1 and insulating
when $U/W >$1. A transition from non-magnetic metal to an antiferomagnetic insulator is thus expected to occur as
$U/W$ is varied within this picture of distinction. V$_2$O$_3$ is the most widely studied example with a low
temperature antiferromagnetic insulating phase which transforms into a high-temperature paramagnetic metal.
However, studies over the last decade have identified several strongly correlated systems which exhibit an
antiferromagnetic metal ground state. These systems seem to be characterized by strong hybridization of the metal
$3d$ and ligand $2p$ states, leading to an intermediate $U/W$. Examples include
(La$_{1-z}$Nd$_z$)$_{1-x}$Sr$_x$MnO$_3$ \cite{aki}, Pr$_{0.5}$Sr$_{0.5}$MnO$_3$ \cite{kaji}, CrN \cite{pab},
CrB$_2$ \cite{gre}, CaFe$_2$As$_2$ \cite{dia}, BaFe$_2$As$_2$ \cite{ana} etc.

Yet another exciting compound possibly belonging to this family is the Cr$^{4+}$ based perovskite: CaCrO$_3$. This GdFeO$_3$ type distorted pervoskite has an orthorhombic structure with \textit{Pbnm} space group, and orders antiferromagnetically at T$_N \sim$ 90 K (Ref. \cite{khom2}). The magnetic structure is $C$-type, i.e. ferromagnetic chains along $c$ direction stacked antiferromagnetically, with a magnetic moment of 1.2 $\mu_B$, as determined from neutron diffraction \cite{khom2}. Moreover, neutron diffraction study also revealed the pronounced structural distortion taking place at T$_N$ with contraction along the $c$-axis and expansion along $a$- and $b$-axis. The electronic nature of this material is quite ambiguous, as reflected by its electrical transport study. Based on the observation of rise in electrical resistivity with fall in temperature, polycrystalline CaCrO$_3$ was first proposed to be a semiconductor \cite{good-mat}. Subsequent report on single crystals showed a decrease in resistivity with the fall in temperature stating it to be a metal \cite{wei}. Recently, a pressure dependent electrical resistivity study of CaCrO$_3$ and SrCrO$_3$ proposed that these compounds are semiconductors at ambient pressure, which undergo a transition to a metal with increase in pressure \cite{zhou}. On the other hand, a subsequent study using infrared reflectivity and optical conductivity measurements point to a three-dimensional metallic nature and highlights the importance of correlation effects, stating CaCrO$_3$ to be itinerant but close to being localized \cite{khom2}. Efforts to calculate the electronic density of states (DOS) of CaCrO$_3$ have been made using the local spin-density approximation (LSDA) and LSDA+U (U: on-site Coulomb correlations) \cite{khom1}. A metallic behavior with Stoner-like band magnetism is found using the LSDA approach. Whereas, LSDA+U results in an insulating ground state with (i) a small gap of 0.05 eV for calculations that adhere to the actual crystal symmetry and (ii) a robust gap of 0.5 eV for a symmetry unrestricted approach, indicating that the calculated DOS for CaCrO$_3$ is quite sensitive to its crystal symmetry \cite{khom1}. The magnetic order in this case results from a superexchange interaction of localized electrons. It is evident that the key to resolve the ambiguity raised over the electronic ground state of CaCrO$_3$ lies in understanding the character of $d$ electrons and the possible role of Coulomb correlations. Therefore, a knowledge of the electronic structure of CaCrO$_3$ using spectroscopic experiments that reflect the Cr $3d$ states is essential.

The present study thus aims to address the nature of the electronic structure and identify the role of correlations in CaCrO$_3$ using soft x-rays and laser photoemission spectroscopy (PES). We establish that CaCrO$_3$ is an antiferromagnetic metal with large on-site electron correlation energy, $U$. We also show that it is the sizable hybridization between Cr $3d$ and O $2p$ that prevents CaCrO$_3$ from being an insulator inspite of the strong correlations. In particular, we have carried out resonant PES across Cr $2p \to 3d$ threshold, the valence band (VB), Cr $2p$, Ca $2p$ and O $1s$ core-level spectra at 25 K i.e. in the antiferromagnetically ordered state. Using synchrotron radiation, measurement of resonant PES with excitation energy tuned across the transition-metal $2p-3d$ threshold, is a useful technique to reveal the transition-metal $3d$ electron character in the total VB, as it provides site-projected and orbital-selective $3d$ partial DOS \cite{allen}. Further, while the metallicity in CaCrO$_3$ is confirmed by its valence band and core-level spectra, resonant PES shows the presence of strong Coulomb correlations in CaCrO$_3$. In addition, X-ray absorption spectroscopy (XAS) has been carried out in the high temperature paramagnetic phase at 120 K and the low temperature antiferromagnetically ordered phase at 25 K, across the Cr $L_{3,2}$ edge and O $K$-edge. XAS shows small temperature dependent changes across T$_N$, thus revealing the effect of structural distortion in the form of small changes in the Cr $3d$ - O $2p$ hybridization.

\section*{Experimental Details}
Polycrystalline CaCrO$_3$ used in the present study was prepared by high pressure synthesis and thoroughly characterized for its transport and magnetic properties \cite{khom2}. Soft X-ray photoemission and XAS experiments were performed at BL17SU SPring-8. The data was obtained using Gammadata-Scienta SES2002 electron analyzer. The total energy resolution for PES was $\sim$ 200 meV and the data was collected in the angle-integrated mode. Resonant PES spectra were normalized to scan time and incident photon flux. XAS was recorded in the total electron yield mode. Laser PES was performed using a Scienta R4000WAL electron analyzer and a vacuum-ultraviolet laser (h$\nu$= 6.994 eV) \cite{kiss} and the total energy resolution was set to 5 meV at 10 K. All the measurements were carried out on a clean sample surface obtained by fracturing the sample \textit{in situ} at the lowest temperature of measurement. A vacuum below 4 x 10$^{-8}$ Pa was maintained throughout the measurements. All the spectra were calibrated using the Fermi level obtained for a gold film evaporated onto the sample holder.

\section*{Results and Discussion}
\begin{figure}
\centering
\includegraphics[width=1\columnwidth]{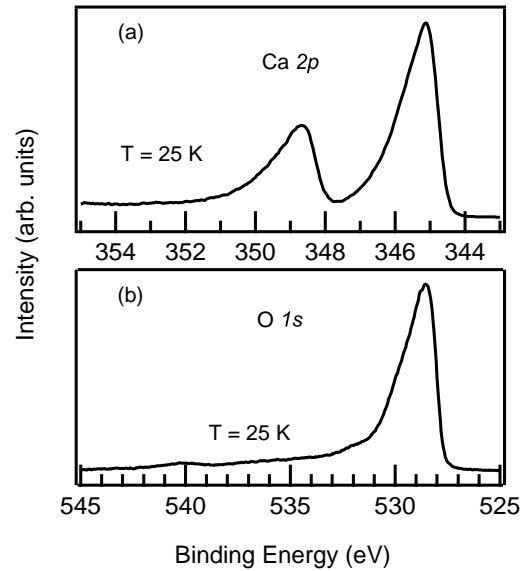}
\caption{\label{CaO}(a) Ca $2p$ and (b) O $1s$ core-level spectra obtained using 1200 eV excitation energy.}
\end{figure}
Figure \ref{CaO}(a) and (b) presents the Ca $2p$ and O $1s$ core-level spectra obtained using 1200 eV incident
photon energy. Sharp and clean spectra with no spurious features at higher binding energy (BE) side of the main
peak confirm high quality of the sample surface. The O $1s$ has a BE of 528.5$\pm$0.2 eV that matches with the
value of 529.3 eV reported for another Cr$^{4+}$ compound, CrO$_2$ \cite{jcc}. The fairly broad spectrum however
hints towards more than one component contributing towards the total intensity of O $1s$ signal. It is well
established from the neutron and x-ray powder diffraction studies of CaCrO$_3$ \cite{khom2, zhou} that the room
temperature \textit{Pbnm} structure undergoes distortion simulatenous to the magnetic ordering at T$_N \sim$ 90
K. As a result, the lattice parameter along the $c$-axis shrinks, while that along $a$ and $b$ elongate, yielding
three different Cr-O bond distances; with the apical bond Cr-O1 = 1.8954(5)\AA,~ and the in-plane bonds Cr-O2 =
1.901(6)\AA~, \& 1.919(5)\AA, (as stated in Ref.\cite{zhou}). We thus have three different types of oxygen ions
contributing towards the total O $1s$ spectrum that can result in one broad signal. The Ca $2p$ spectrum exhibits
a BE typical of Ca$^{2+}$ ($\equiv$ 345 eV), but with an overall asymmetric (Doniach-\v{S}unji\'{c}) line shape,
an outcome of electron-hole pair shake-up process that is characteristic of a metal. We confirm the metallic
behaviour from the VB measurements in the following.

\begin{figure}
\centering
\includegraphics[width=1\columnwidth]{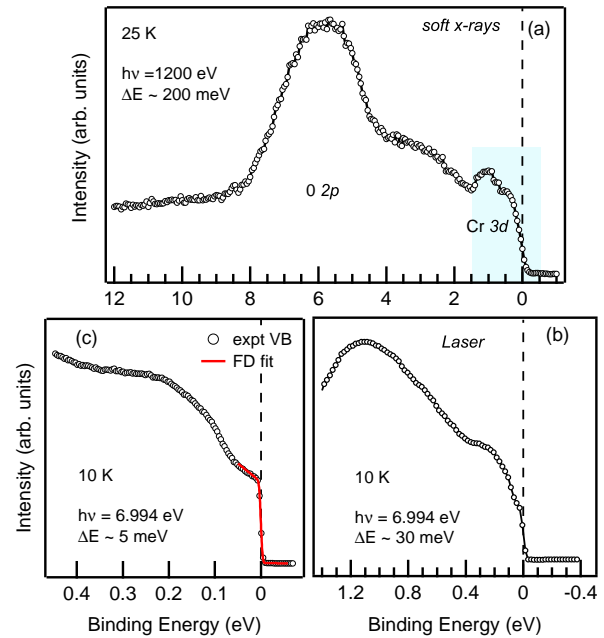}
\caption{\label{vb}(Color online)(a) Soft x-ray VB spectrum of CaCrO$_3$ obtained at 25 K. A clear metallic Fermi edge can be seen. The shaded portion was scanned using laser source and the corresponding VB spectrum is shown in (b). The feature at 0.2 eV is clearly seen. (c) Resolution-convoluted Fermi Dirac function (solid line) fit to the experimental VB.}
\end{figure}

The VB spectrum of CaCrO$_3$ in the antiferromagnetically ordered phase is presented in Fig. \ref{vb}(a). Clear signature of metallicity is observed as the spectral intensity crosses the Fermi edge (E$_F$) indicating presence of finite density of states (DOS) at E$_F$. This is an interesting result as the commonly known metallic oxides are either correlated-paramagnets (ex: LaNiO$_3$ \cite{shre}) or ferromagnets (ex: CrO$_2$ \cite{gro}, SrRuO$_3$ \cite{kli}). Besides, though a rise in electrical resistivity was observed with the lowering of temperature in polycrystalline CaCrO$_3$, an activation energy could not be extracted \cite{good-mat,zhou,khom2}. Infact, the metallic nature of antiferromagnetic phase of CaCrO$_3$ was also deduced from its infra-red reflectivity and optical conductivity measurements \cite{khom2}. Comparing the experimental VB with the reported band-structure calculations (LSDA+U approach) \cite{khom2, khom1}, the peak centered at $\sim$ 1.2 eV below E$_F$ can be attributed to the Cr $3d$ states. While, the broad band centered at $\sim$ 6 eV below E$_F$ comprises dominantly of the O $2p$ states.

In order to ascertain the low energy scale electronic structure of CaCrO$_3$ in the vicinity of E$_F$, we measured bulk sensitive laser PES at 10 K using an excitation energy of h$\nu$=6.994 eV and 5 meV resolution (see Fig. \ref{vb}(b)\&(c)). Apart from the sharp Fermi edge that reiterates the metallic nature of CaCrO$_3$, laser VB depicts a clear hump around 0.2 eV below E$_F$. Thus it appears that the spectral weight of Cr $3d$ band dominant at E$_F$ is distributed in peaks centered at $\sim$ 1.2 and $\sim$ 0.2 eV BE. Similar features have been reported in the VB of early transition metal oxides such as Ca$_{1-x}$Sr$_x$VO$_3$, Y$_{1-x}$Ca$_x$TiO$_3$, leading to their classification as Mott-Hubbard correlated metals \cite{mori1, ino, mori2}. In these materials the feature very near to E$_F$ has been attributed to the itinerant $d$-band states or coherent part of the spectral function, while that present at higher BE has been assigned to the incoherent part or the lower Hubbard band and corresponds to electronic states localized due to electron-electron correlations. Additionally, LSDA calculations \cite{khom2, khom1} predict the Cr $3d$ states to occur at $<$ 1 V BE, while it is evident from Fig. \ref{vb} that the bandwidth of Cr $3d$ is spread over 2 eV. Hence, the experimental VB spectrum obtained here suggests a picture that goes beyond the LSDA calculations.

In light of such reasoning, inclusion of electron-electron correlation effect in band structure calculations seems imperative for the description of electronic structure of CaCrO$_3$. In this regard, the LSDA+U calculations reported so far \cite{khom1} show that the calculated DOS for CaCrO$_3$ is quite sensitive to its crystal structure. In their report, Streltsov \cite{khom1} \textit{et.al} state that the total DOS obtained with the full set of the \textit{Pbnm} crystal symmetry operations results in a small gap of 0.05 eV and slight changes in the basis set or calculation parameters like $U$ result in closing of the gap. On the other hand, an insulating ground state with a gap of 0.5 eV and orbital ordering is observed for a symmetry unrestricted calculation. This band gap is very robust and persists even if $U$ is decreased down to 1.8 eV. Irrespective of the choice of solution based on crystal symmetry, it is found that the introduction of on-site Coulomb correlations leads to a strong mixing of occupied Cr $3d$ and O $2p$ states. The same study also noted that the magnetic interactions are governed by sizable $pd$ hybridization. Having observed a clear metallic Fermi edge, the experimental data suggests that the LSDA+U results obtained using full crystal symmetry provides a better description of the experimental VB spectrum. However, it becomes important to obtain an estimate of $U$ from the experimental data. Accordingly, resonant VB spectra were obtained by tuning the excitation energy across the Cr $2p-3d$ absorption, which significantly increases the relative photoemission cross-section of Cr $3d$ states over O $2p$ states. In order to obtain the resonant PES, we first measured the XAS across the Cr $L_{3,2}$-edge. We have also carried out measurements at the O $K$-edge above and below T$_N$.

\begin{figure}
\centering
\includegraphics[width=1\columnwidth]{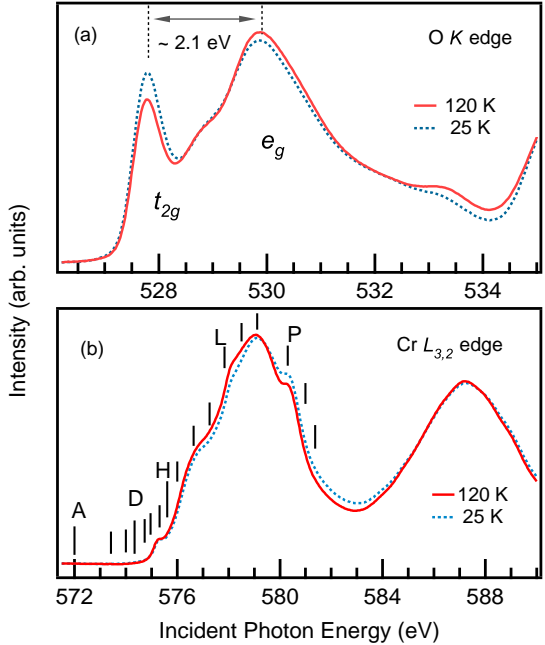}
\caption{\label{xas}(Color online) XAS measured as a function of temperature above and below antiferromagnetic ordering temperature T$_N$ = 90 K. (a)O $K$-edge and (b)Cr $L_{3,2}$ edge. Temperature dependent changes are seen in both the spectra.}
\end{figure}

Figure \ref{xas}(a) presents the O $K$-edge XAS that originates from transition to unoccupied O $2p$ states hybridized with the metal states and hence reflects features of metal $3d$ character \cite{tana}. Given that Cr$^{4+}$ in CaCrO$_3$ has a $d^2$ configuration, the first peak around 528 eV can be assigned to the relatively empty $t_{2g}$ band. Above the $t_{2g}$, one can see an empty $e_g$ band centered at $\sim$ 530 eV. An energy separation of $\sim$ 2.1 eV between these two peaks is an estimate of the crystal-field parameter, 10D$q$.
Fig. \ref{xas}(a) also shows the temperature-dependence of O $K$-edge where an increase in the relative intensity of the first peak is seen at low temperature. Such an increase in the first peak intensity hints towards a small decrease of O $2p$ - Cr $3d$ hybridization in the low temperature phase. A small but definite temperature dependent change is simultaneously observed in the Cr $L_{3,2}$-edge XAS as seen in Fig. \ref{xas}(b). While the spectrum for Cr $L_{3,2}$-edge consists of many multiplets, certain features (especially those indicated with `L' and `P') display clear change in spectral weight on decreasing temperature. The difference in the apical and in-plane Cr-O bond-distances known from low temperature crystal structure studies \cite{khom2, zhou} are considered responsible for such a spectral weight transfer.

\begin{figure}
\centering
\includegraphics[width=1\columnwidth]{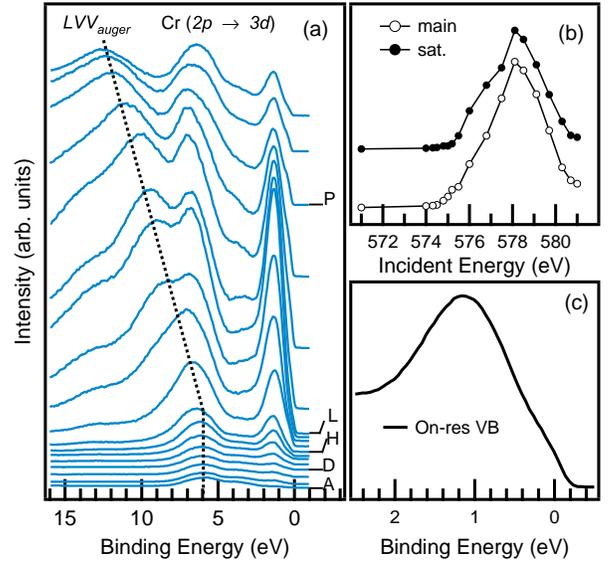}
\caption{\label{cis}(Color online)(a) The resonant PES VB spectra recorded using excitation photon energy across Cr $2p-3d$ transition. Dotted line shows the evolution of the satellite feature into Auger feature. (b) CIS for the main Cr $3d$ peak at 1.2 eV and the satellite feature at 5.9 eV, normalized to the intensity at 571 eV and have arbitrary offset. (c) On-resonance VB reiterates the two-peak structure of Cr dominated states at $E_F$ with features at 1.2 and 0.2 eV.}
\end{figure}
Next, we consider the presence of electron-electron correlations in CaCrO$_3$. The resonant VB spectra shown in Fig. \ref{cis}(a) were recorded with excitation energy ranging from 571 to 581 eV. The actual incident photon energies are indicated by tick marks on the Cr $L_{3}$ XAS curve presented in Fig. \ref{xas}(b). The relative photoionization cross section of Cr $3d$ to O $2p$ being significantly increased, the 1.2 eV peak is enhanced with the increasing photon energy. Further, an Auger signal ($LVV_{auger}$) with an apparent binding energy of about 11.7 eV for the maximum incident photon energy spectrum (topmost curve) can be easily identified by its dependence on the excitation energy. The Auger feature converges towards the satellite peak at $\sim$ 5.9 eV for the lower incident photon energy (dotted line shows the convergence of Auger peak). Constant-initial-state (CIS) spectra of the main peak (1.2 eV) and the Auger peak are presented in Fig. \ref{cis}(b). These spectra follow the same photon energy dependence as Cr $L_{3}$ curve, thus indicating that the Cr $3d$ weight is indeed spread over a wide energy range. It also serves to confirm the Cr $3d$ character of the so called two-hole correlation satellite \cite{huf}. These observations indicate the presence of strong on-site Coulomb interaction energy ($U$) in CaCrO$_3$, as has been demonstrated for Cr metal\cite{huf}, Cr silicide\cite{galan} and Cr nitride \cite{pab}. Besides, the VB spectrum obtained at excitation energy equal to the maximum of Cr $L_{3}$ XAS i.e. the \textit{on-res} VB spectrum, shown in Fig. \ref{cis}(c) emulates the laser VB spectrum with a two peak structure of Cr $3d$ states with a main Cr $3d$ peak at 1.2 eV and a weak shoulder at $\sim$0.2 eV.

\begin{figure}
\centering
\includegraphics[width=1\columnwidth]{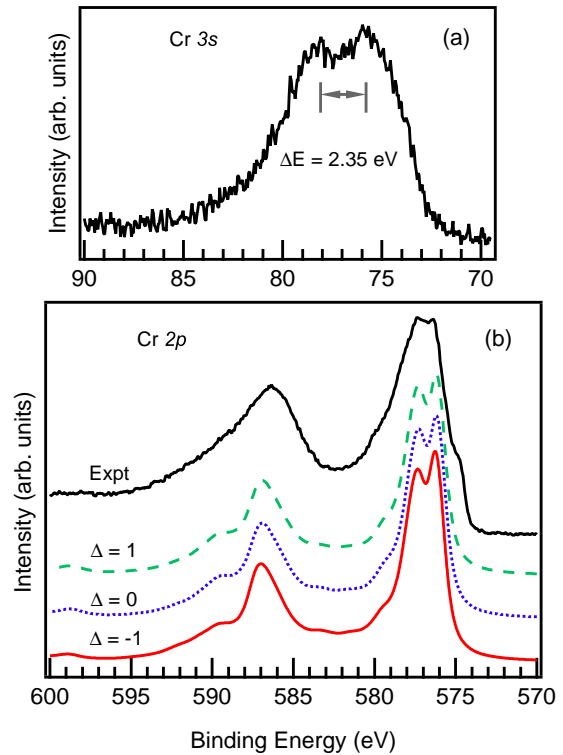}
\caption{\label{core}(Color online) Core level spectra of Cr (a) $3s$ and (b) $2p$. Multiplet features are present in both the spectra. The calculated spectra for Cr $2p$ using cluster model calculations including metallic screening channel at $E_F$ are compared with the experimental spectrum (\textit{See text for more details}).}
\end{figure}

By employing an analysis similar to that of CrN \cite{pab}, we obtain an estimate of electron-electron correlation energy in CaCrO$_3$. The expression to obtain $U$ can be stated as, $$U = E_{2p} - (h\nu - BE_{auger}) - 2\epsilon_{3d}$$ where, $E_{2p}$ is the binding energy of Cr $2p$, $h\nu$ is the incident photon energy and $BE_{auger}$ gives the BE of the corresponding Auger peak. $2\epsilon_{3d}$ is the average energy of two uncorrelated holes, obtained from the self convolution of the one-electron removal Cr $3d$ peak occurring at 1.2 eV. On substituting the experimental values of $E_{2p}$ = 576.5 eV, obtained as the BE of main peak in the Cr $2p$ spectrum (Fig. \ref{core}(b)), $(h\nu - BE_{auger})$ = 569.33 eV and $2\epsilon_{3d} \sim$ 2.4 eV, we obtain an estimate of $U\sim$ 4.8 eV for CaCrO$_3$. Thus summing up the experimental results presented so far, we establish that CaCrO$_3$ is an antiferromagnetic metal with large on-site electron correlation energy, $U$.

Finally, the Cr $3s$ and $2p$ core-level spectra of CaCrO$_3$ at T = 25 K is presented in Fig. \ref{core}(a)\&(b) respectively. The $3s$ spectrum exhibits a doublet with energy separation of about 2.35 eV. Such multiplet splitting of $3s$ core-level in transition metal compounds primarily originates from the exchange coupling between the $s$ core hole spin and the $3d$ spin. It is expressed as $\Delta E = (2S+1)J_{sd}$ where $S$ is the magnitude of total spin of the $3d$ electrons and $J_{sd}$ denotes the effective exchange integral between the $3s$ core hole and the $3d$ electron. The splitting of $3s$ spectrum hence reflects a well defined local magnetic moment of the corresponding transition metal ion \cite{fad}. Another Cr$^{4+}$ compound, CrO$_2$ that has a local magnetic moment of 2$\mu_B$, shows a slightly larger $3s$ splitting of 3.5 eV \cite{jcc}.

The Cr $2p$ spectrum presented in Fig. \ref{core}(b) also shows multiplet features besides the $2p_{3/2}$, $2p_{1/2}$ spin-orbit splitting. The BE of the main features are $\sim$ 577.3 eV and 576.5 eV. The multiplet splitting of Cr $2p_{3/2}$ has also been reported for insulating compounds like Cr based sulphides and selenides where it was conjectured to be due to the local magnetic moment on Cr ions, with a moment of 2.9 - 3.0 $\mu_B$ resulting in a splitting of 0.95 - 1.0 eV \cite{tsur}. In addition a well-screened feature is observed at 575 eV, on the low BE side of the main peak in the $2p$ core-level spectrum. The occurrence of such a feature in transition metal compounds has been attributed to a screening channel derived from the states at $E_F$ responsible for metallicity \cite{tagu}. In very simple terms, the well-screened feature develops due to the core-hole potential in the photoemission final state being screened by the charge transfer from the conduction band states at $E_F$.

With the aim of accounting for the origin of features present in the core-level spectrum of CaCrO$_3$, we have carried out cluster model calculations. In addition to the usual calculations based on a CrO$_6$ cluster, we introduced a charge transfer from a coherent band, $C$, at E$_F$. As shown originally in Ref. \cite{kot13}, such a charge transfer from the coherent states can be directly related to the metallic screening in the core-level PES. Analogous to the parameters associated with Cr $3d$ and O $2p$ ligand states viz. charge transfer energy $\Delta$ and hybridization $V$, the additional parameters were $\Delta^*$ - the charge transfer energy between Cr $3d$ and $C$, and $V^*$ - the hybridization between Cr $3d$ and $C$. The other parameters involved in the calculations were the on-site Coulomb interaction energy $U$($\equiv$ 4.5 eV, obtained from the resonant PES above), the crystal field splitting $10Dq$($\equiv$ 2.1 eV, obtained from O $K$-edge XAS), Coulomb interaction between Cr $3d$ and Cr $2p$ core hole states $U_{dc}$ (typically set to 1.2 - 1.4$U$). Under these constraints we varied $V$, $V^*$, $\Delta$ and $\Delta^*$, and found satisfactory results for $V$ = 0.6 eV, $V^*$ = 2.0 eV with $\Delta$ and $\Delta^* \sim$ 0. Since a recent LSDA+U study proposed a negative $\Delta$ as the possible ground state for CaCrO$_3$, \cite{khom2, khom1} we further varied $\Delta$ from -1, 0, to 1 eV to check for this possibility.
\begin{figure}
\centering
\includegraphics[width=1\columnwidth]{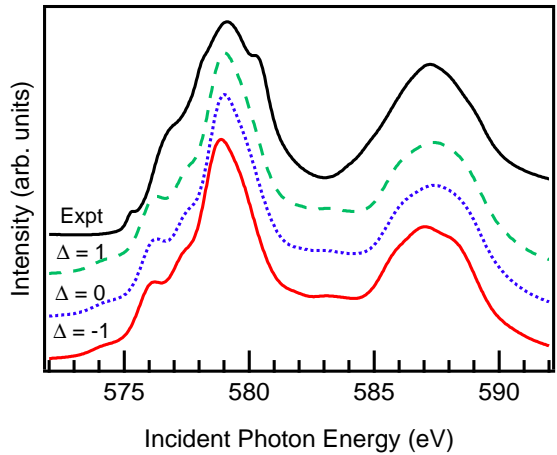}
\caption{\label{cal}(Color online) Cr $L_{3,2}$-edge XAS spectrum compared with the calculated spectra obtained using cluster model calculations with inclusion of metallic screening channel at $E_F$.}
\end{figure}

Figure \ref{core}(b) \& \ref{cal} contain the calculated spectra together with the experimentally obtained PES and XAS spectra respectively. As can be seen in Fig. \ref{core}(b) for the PES results, the calculated spectra show a splitting of the main peak as seen in the experimental spectra but the low energy well-screened feature overlaps the main peak and is not suitably reproduced. The change in $\Delta$ also leads to negligible effects on the spectra and all the three cases seem equally likely. However, on comparing the experimental and calculated spectra of Cr $L_{3,2}$-edge XAS in Fig. \ref{cal}, the spectrum with $\Delta$ = -1 eV provides a better fit particularly for the $L_2$-edge with negligible differences for the $L_3$ region. The results suggest a small negative charge transfer energy for the ground state of CaCrO$_3$. While further studies could provide a more definitive picture, for a system like Cr$^{4+}$ it is reasonable to expect a small or negative charge transfer energy. This is particularly true in the presence of strong hybridization as is also known for other systems like NaCuO$_2$ \cite{mizo}, CrO$_2$ \cite{koro}, CuFeS$_2$ \cite{kat}.

\section*{Conclusion}
To summarize, our photoemission spectroscopy measurements on CaCrO$_3$ clear the ambiguity cast over its electronic structure from electrical transport measurements. The valence band spectrum shows a clear Fermi edge, confirming the metallic nature of CaCrO$_3$ in the low temperature phase. In the vicinity of E$_F$, the Cr $3d$ dominated band exhibits a two-peak feature attributable to the localized and itinerant $d$-band states, quite like the Mott-Hubbard type systems. Analysis of resonant PES across the Cr $2p\to3d$ threshold yields an on-site Coulomb energy $U \sim$ 4.8 eV, indicating presence of strong electron-electron correlations. Inspite of the large $U$, the system is metallic in the antiferromagnetically ordered low temperature phase. The small temperature dependent changes observed in Cr L$_{3,2}$ and O $K$ edge XAS are consistent with a small change in the Cr $3d$- O $2p$ hybridization. The main features of the Cr $2p$ PES and the $L_{3,2}$-edge XAS can be reproduced by model cluster calculations which suggest a small negative charge transfer energy between Cr $3d$ and O $2p$ states. The results indicate that CaCrO$_3$ can be best described as a strongly hybridized antiferromagnetic metal exhibiting a mixed Mott-Hubbard and Charge-transfer character.

\begin{acknowledgments}
P. A. Bhobe acknowledges support from Japan Society for Promotion of Science. The synchrotron radiation experiments were performed at BL17SU, SPring-8 with the approval of RIKEN (Proposal No. 20091140).
\end{acknowledgments}

\end{document}